# Anisotropic Field Suppression of Morin Transition Temperature in Epitaxially Grown Hematite Thin Films

Haoyu Liu[1], Hantao Zhang[2], Josiah Keagy[1], Qinwu Gao[3], Letian Li[3], Junxue Li[3], Ran Cheng[2,1], and Jing Shi[1]

1. Department of Physics & Astronomy, University of California, Riverside, CA 92521, USA
2. Department of Electrical & Computer Engineering, University of California, Riverside CA 92521, USA
3. Department of Physics, Southern University of Science and Technology, Shenzhen 518055, China

We have demonstrated the existence of the Morin transition in epitaxially grown hematite thin films exceeding a critical thickness. The Morin transition temperature can be suppressed by magnetic fields applied both parallel and perpendicular to the Dzyaloshinskii-Moriya (DM) vector, exhibiting a distinct anisotropic behavior that is consistent with bulk hematite crystals. Detailed analysis explains the anisotropic behavior and provides a method for determining the DM strength, which remains nearly constant across the sample thickness over four orders of magnitude. Our findings obtained with transport measurements offer a valuable approach for studying antiferromagnetic spin configurations in thin films and nanodevices.

Hematite (α-Fe$_2$O$_3$) has emerged as a promising material for antiferromagnetic spintronics [1–5] due to its unique properties. Unlike other antiferromagnets, hematite exhibits a Morin transition [6], a phase change from the easy-plane to the easy-axis configuration as temperature $T$ is lowered. This transition is accompanied by a reorientation of the Néel vector, the antiferromagnetic order parameter. In bulk crystals, the Morin transition temperature ($T_M$) is typically around 260 K.

Another distinctive feature of hematite is the presence of the Dzyaloshinskii-Moriya (DM) interaction with the DM vector along the c-axis [7–9], owing to its crystal symmetry. In the easy-plane phase ($T > T_M$), this interaction results in canted moments in the hexagonal basal plane, leading to weak ferromagnetism. The DM interaction enables various phenomena that are highly useful for spin-torque switching [10–13], Néel vector detection [14–17], and microwave coupling [18,19]. However, experimentally determining the DM interaction strength in hematite remains a challenging task [9,20,21].

Thin films are essential for antiferromagnetic spintronics applications. In hematite thin films, the Morin transition temperature has been observed to vary widely, ranging from zero to room temperature [10,17,22–26]. This variability is likely influenced by interfacial strain-induced anisotropy, similar to what is observed in ferromagnetic films; however, the limited exploration has not led to a clear consensus [22,27–29]. At this point, it is still unclear whether the DM interaction strength varies in thin films and to what extent it contributes to the observed $T_M$ variation.

In this Letter, we first investigate the epitaxial growth of hematite thin films using pulsed laser deposition (PLD). Given the antisymmetric nature of the DM interaction, we anticipate anisotropic spin reorientation responses in hematite when a magnetic field ***H*** is applied parallel or perpendicular to the DM vector. To study the Néel vector reorientation in thin films, we perform transport measurements utilizing the spin Hall magnetoresistance (SMR) effect in hematite/Pt heterostructures.

High-quality α-Fe$_2$O$_3$ thin films are epitaxially grown on Al$_2$O$_3$(0001) substrates using PLD (see Supplemental Material). To investigate the presence of the Morin transition in α-Fe$_2$O$_3$ thin films, various growth conditions have been explored. An optimized growth recipe is established, beginning with a substrate temperature of 700 °C, which is then reduced to a range between 300 °C and 700 °C once the seeding layer reaches a thickness of approximately 1 nm. The 2θ/ω x-ray diffraction (XRD) results in Figure 1(a) exhibit (0006) peaks, indicating that the α-Fe$_2$O$_3$ thin films of various thicknesses (~9 nm, ~60 nm, and ~150 nm) are (0001)-oriented with excellent quality. The inset displays a magnified view of the (0006) peaks for 60-nm-thick and 150-nm-thick samples, clearly showing the Kiessig fringes. Real-time reflection high-energy electron diffraction (RHEED) monitoring reveals streaky patterns throughout the growth process as represented by the inset of Figure 1(b). Atomic force microscopy confirms the formation of an atomically smooth surface with distinct terraces, exhibiting a root-mean-square (RMS) roughness of approximately 0.1 nm (Figure 1(b)). Additionally, more structural characterization data including rocking curves (Figure S1b, right), pole figures (Figure S2a, left) and $\phi$-scan data (Figure S2b, right) are shown in the Supplemental Material. These results demonstrate high crystalline quality that is better than or comparable with that of the existing hematite films [10,11,17,22–29].

The representative sample discussed in the main text is grown at 500 °C. Using the optimized growth condition discussed above, we observe the Morin transition in thin films with thickness exceeding 40 nm. As depicted in Figure 1(c), the magnetometry results of a 180 nm thick film at zero magnetic field clearly demonstrate the Morin transition. Around 280 K, the spins reorient from an alignment along the c-axis to the basal hexagonal plane (blue), resulting in an abrupt jump in magnetization. This sharp change indicates the onset of the easy-plane phase, characterized by canted moments arising from the DM interaction. The negative magnetization below $T_M$ is from the diamagnetic substrate Al$_2$O$_3$ and sample holder. We hypothesize that the thickness dependent Morin transition is influenced by variations in lattice mismatch [22]. When the out-of-plane lattice strain is compressive (thickness < 40 nm), the Morin transition does not occur. However, when it is tensile (thickness > 40 nm), the Morin transition is present, as illustrated in Figure 1(d). The lattice strain comes from the interplay of growth temperature [27] and thickness dependence of epitaxial strain from the sapphire substrate. We believe that because of the different thermal expansion coefficient of hematite from the clamping substrate, growth and annealing temperature could be effectively employed to tune the strain of the epitaxial hematite thin films.

While magnetometry is a valuable tool for studying the magnetic properties of hematite, its limited sensitivity to small canted moments and the Néel vector orientation hinders its application to antiferromagnetic thin films. As in insulating ferromagnets [30], the spin configurations of insulating antiferromagnets can be probed resistively using the proximity effect [31–33] or SMR [23,27,29,34–38]. This involves incorporating a thin heavy metal layer such as Pt in close contact with the insulator to indirectly sense the antiferromagnetic order. Thus, to electrically detect the Morin transition, we fabricate a 250-μm-wide Hall cross using standard photolithography. A 3-nm-thick Pt film is then deposited onto the α-Fe$_2$O$_3$ thin films via DC magnetron sputtering. An AC current of 1 mA is applied through the Pt channel, while the longitudinal and transverse voltages are simultaneously measured using a multimeter and a lock-in amplifier.

Figure 2(a) presents the angular dependence of the transverse voltage $V_{xy}$ for an α-Fe$_2$O$_3$(60 nm)/Pt bilayer under different applied magnetic fields rotating in the basal plane below and above the Morin temperature. At room temperature (301 K), the $V_{xy}$ oscillations show two-fold symmetry with the first maximum located at 45°. Considering the expected 45° phase shift with respective to the $V_{xx}$ oscillations as shown in Figure S42, observed $V_{xy}$ results are consistent with the characteristic negative SMR behavior in α-Fe$_2$O$_3$ as previously reported in easy-plane antiferromagnet/heavy metal heterostructures [10,12,34,35], suggesting a rotating Néel vector-related SMR. For a low applied field (around $\mu_0H$ ~ 0.1 T), hysteresis is evident in the round-trip measurement, indicating the presence of a finite magnetic anisotropy in the film plane [39]. As the field increases to 0.5 T, the Néel vector rotates freely with the field, and hysteresis is no longer discernible. Consequently, the SMR signal can be fitted by a sine-squared function. The amplitude of this function saturates quickly at low fields and remains nearly constant up to $\mu_0H$ ~14 T [34,35]. A slight decrease at higher fields can be attributed to the increased canted moments induced by stronger applied fields, which oppose the SMR caused by the Néel vector contribution [40]. In contrast, at 2 K, the in-plane rotation data retains a sine-squared shape and phase similar to 300 K but exhibits a less steep, monotonically increasing amplitude as the field increases. The distinct behavior observed at 2 K indicates a transition to the easy-axis phase from the easy-plane phase at 300 K. The linear trend reflects the characteristic smooth Néel vector rotation in the easy-axis phase [23,36,41], which will be discussed in more detail later. To further confirm the presence of the easy-axis phase, we perform SMR measurements under sweeping magnetic fields below 300 K. With the field applied along the c-axis, the spin-flop transition occurs at a critical field, resulting in a sudden jump in the transverse voltage $V_{xy}$ as shown in Figure 2(c). This SMR jump arises from the abrupt Néel vector reorientation. The spin-flop field gradually decreases from $\mu_0H$ ~ 7 T at low temperatures to zero as the discontinuity in the transverse voltage vanishes at 280 K, which corresponds to the Morin temperature of this hematite thin film.

After demonstrating the remarkable sensitivity of SMR to spin-flop transition in hematite films, we employ this sensitive probe to investigate the Morin temperature where the spin reorientation occurs. By applying a magnetic field parallel and perpendicular to the c-axis (the DM vector direction), we compare the field responses between these two field orientations. The upper panel of Figure 3(a) illustrates the temperature dependence of transverse voltage responding to $H$ applied along the c-axis ($H$ // $c$), i.e., parallel to the DM vector. There is an obvious jump in $V_{xy}$ signal. A clear shift of the $V_{xy}$ jump to lower temperatures is observed with increasing magnetic field strength. This trend indicates that the application of a magnetic field along the DM vector stabilizes the easy-plane phase, thereby suppressing the Morin transition temperature $T_M$. Interestingly, when $H$ is applied perpendicular to the c-axis ($H \perp c$), $T_M$ is also suppressed to lower temperatures, as shown in the lower panel of Figure 3(a). However, the field response of $T_M$ for $H \perp c$ differs qualitatively from that for $H$ // $c$, indicating a strong anisotropy in the field-induced Morin transition suppression. Despite the differing trends, the jump becomes less pronounced as the magnetic field strength increases for $H \perp c$, contrasting the behavior observed for $H$ // $c$. Figure 3(b) summarizes the anisotropic field responses of $T_{M_s}$, which is defined as the onset of the transition (see Supplemental Material S2 for details) The field dependence data are fitted and shown by solid lines. For $H$ // $c$, $T_M$ decreases quadratically with the field, whereas for $H \perp c$, it decreases linearly. The two curves intersect at $\mu_0 H$ ~3.8 T, above which $T_M$ decreases much more steeply for $H$ // $c$. These results are reproducible in thicker films with similar cross points.

To further validate the anisotropic field suppression phenomenon in thin films, we perform both magnetization and transport measurements on a bulk α-$Fe_2O_3$ crystal. Similar qualitative anisotropic responses are observed in both quantities, with the magnetization responses illustrated in Figure 4(a). The contrast in $T_M$ suppression between the $H$ // $c$ and $H \perp c$ directions, as extracted from the magnetometry measurements, is clearly presented in Figure 4(b). The fitted quadratic and linear curves intersect at approximately $\mu_0 H$ ~3.6 T. Detailed transport measurements are provided in the Supplemental Material (Figure S3 S4 and S5).

Strikingly, this anisotropic behavior is invariant with respect to hematite thickness. Whether in thin films of tens-of-nm thick or bulk crystal of few-mm large, applying magnetic fields parallel or perpendicular to the c-axis consistently leads to distinct responses and a cross point near $\mu_0 H$ ~3.6 T. Furthermore, such an anisotropic response is somewhat unexpected. One might intuitively expect a stronger external field applied along the c-axis ($H \parallel c$) should favor the spin-flop configuration by destabilizing the easy-axis phase, resulting in a lower $T_M$. Conversely, a field applied perpendicular to the c-axis ($H \perp c$) would not be expected to influence $T_M$, as it is perpendicular to the Néel vector and should not affect its orientation. However, our experimental observations contradict this expectation.

To explain the unexpected $T_M$ decrease for $H \perp c$, we demonstrate the necessity of considering the intrinsic antisymmetric exchange coupling, i.e., the DM interaction, in hematite. We construct a minimal model incorporating essential magnetic interactions while neglecting secondary effects such as misalignment of spins and higher order anisotropies (see Supplemental Material S5). In the presence of the DM interaction, the free energy density $F(\boldsymbol{n})$ can be expressed as

$$F(\boldsymbol{n})/\hbar\gamma = \frac{1}{H_E}[-2H_D \boldsymbol{n} \cdot (\boldsymbol{H} \times \hat{\boldsymbol{z}}) + H_D^2 n_z^2 + (\boldsymbol{H} \cdot \boldsymbol{n})^2] - H_K n_z^2, \quad (1)$$

where $\gamma$ is the gyromagnetic ratio, $n_z$ is the z-component of the normalized Néel vector $\boldsymbol{n}$, $\hat{\boldsymbol{z}}$ is the unit vector along z direction, $H_K$ is the uniaxial anisotropy, $H_E$ is the exchange interaction, and $H_D$ is the DM interaction (all three interactions are in units of kOe) [42–44]. For $\boldsymbol{H} = 0$, $F/\hbar\gamma = \left(-H_K + \frac{H_D^2}{H_E}\right) n_z^2$. The inclusion of the DM interaction modifies $H_K$ by $\frac{H_D^2}{H_E}$, a quantity as small as 0.1 kOe. $T_M$ is then determined by $H_K^{eff}(T_M) = H_K - \frac{H_D^2}{H_E} = 0$, which marks the boundary between the easy-plane phase ($n_z = 0$) and the easy-axis phase ($n_z = \pm 1$). In general, $H_K^{eff}$ is a function of temperature and needs to be determined experimentally. However, without detailed knowledge of $H_K^{eff}(T)$, we make a linear approximation in the vicinity of $T_M$ where $H_K^{eff}$ is small, $H_K^{eff}(T) \approx \alpha(T_M - T)$.

If $H \parallel c$, $F(\boldsymbol{n})$ simplifies to,

$$F(\boldsymbol{n})/\hbar\gamma = \left(-H_K^{eff} + \frac{H^2}{H_E}\right) n_z^2. \quad (2)$$

Solving $F(\boldsymbol{n}) = 0$ yields the field-dependent Morin temperature: $T_\parallel(H) = T_M - \frac{H^2}{H_E \alpha}$, which explains the observed quadratic field dependence of $T_M$ for $H \parallel c$ in Figures 3(b) and 4(b). If $H \perp$ *c*, however, the DM interaction will affect the spin reorientation process, leading to a counterintuitive effect. In this case, $F(\mathbf{n})$ can be expressed as a function of $n_y$,

$$F(\mathbf{n})/\hbar\gamma = \left(H_K - \frac{H_D^2}{H_E}\right)\left(n_y + \frac{HH_D}{H_EH_K-H_D^2}\right)^2 - \frac{H_D^2}{H_E}\left(1 - \frac{H^2}{H_EH_K-H_D^2}\right) - H_K. \tag{3}$$

As the temperature increases across the spin reorientation transition, $n_z$ deviates from $\pm 1$ in the easy-axis phase, accompanied by the emergency of $n_y = \frac{HH_D}{H_EH_K-H_D^2}$ until it reaches -1, or $H_K^{eff} - \frac{H_DH}{H_E} = 0$. This leads to a linear field dependence of $T_M$: $T_\perp(H) = T_M - \frac{H_DH}{H_E\alpha}$. Interestingly, the $T_\parallel(H)$ and $T_\perp(H)$ curves intersect at $H = H_D$, providing a straightforward and effective method for determining the DM interaction strength $H_D$. From Figures 3(b) and 4(b), we estimate the value of $\mu_0H_D$ to be approximately 3.8 T and 3.6 T. The unexpected invariance of the $\mu_0H_D$ value across such a wide range of thicknesses suggests that variations in film properties have minimal influence on the strength of the DM interaction; therefore, the observed DM strength is primarily determined by the intrinsic property of the hematite lattice.

To gain a deeper understanding of the underlying physical mechanism, we numerically simulate the spin reorientation transition by solving the coupled Landau-Lifshitz-Gilbert equations to determine the orientation of $\mathbf{n}$ at various field strengths and temperatures. The simulation results are depicted in Figure 5(a). By carefully selecting the model parameters, we obtain field-dependent responses for both field orientations that quantitatively agree with our experimental data. Notably, when the DM interaction is excluded from the simulations, the field response for $\mathbf{H} \perp \mathbf{c}$ completely disappears, confirming the critical role of the DM interaction in the observed anisotropic field responses of the Morin transition. Additionally, we simulate the spin reorientation transition, as shown in Figure 5(b). The transition exhibits a sharp jump for $\mathbf{H} // \mathbf{c}$ and a smooth crossover for $\mathbf{H} \perp \mathbf{c}$. This behavior can be understood by analyzing the free energy density $F(\mathbf{n})$: for $\mathbf{H} // \mathbf{c}$, the energy ground state from Eq. (2) must be either $n_z = \pm 1$ or 0, depending on the sign of $-H_K^{eff} + \frac{H^2}{H_E}$. For $\mathbf{H} \perp \mathbf{c}$, $n_y$ in the energy ground state described by Eq. (3) ranges from 0 to -1. Correspondingly, $n_z = \pm\sqrt{1-n_y^2}$ gradually changes from $\pm 1$ to 0 during the transition, aligning with the transport results in hematite thin films. Moreover, the crossing point of the simulated curves in Figure 5(b) yields a DM interaction strength $\mu_0H_D$ of approximately 3.3 T, which is in excellent agreement with the experimentally determined value.

Our observed Morin transition in perpendicular fields was previously reported in bulk hematite crystals and identified as a spin-flop transition [41,45]. The field dependence of the Morin temperature essentially mirrors the temperature dependence of the reported spin-flop field. However, it is noteworthy that previous SMR studies on the temperature dependencies of spin-flop fields (Ref. [23] and Ref. [36]) did not reveal a clear anisotropic behavior. Both temperature dependencies were attributed to a scaling with $\sqrt{T - T_M}$.

In summary, we have demonstrated that the Morin transition temperature in hematite is suppressed by the application of external magnetic fields. The field-induced suppression exhibits a strong anisotropy, with a quadratic dependence in one direction and a linear dependence in the orthogonal directions. To our knowledge, such anisotropic field responses in hematite thin films have not been previously reported. We have provided a clear explanation for this anisotropic behavior, attributing it to the DM interaction. By analyzing the field dependence of the Morin transition, we have uniquely determined the strength of the intrinsic DM interaction in thin films, which is consistent with the bulk value. Our findings highlight the critical role of the DM interaction in antiferromagnetic materials.

We would like to thank Yaroslav Tserkovnyak, Derek Reitz, and Igor Barsukov for helpful discussions. H.Y.L., J.K., and J.S. acknowledge the support by NSF-DMR-2203134. H.Z. and R.C. acknowledge the support by Air Force Office of Scientific Research under Grant No. FA9550-19-1-0307. G.Q.W., L.T.L. and J.X.L. acknowledge the support by Guangdong Innovative and Entrepreneurial Research Team Program (Grant No. 2021ZT09C296).

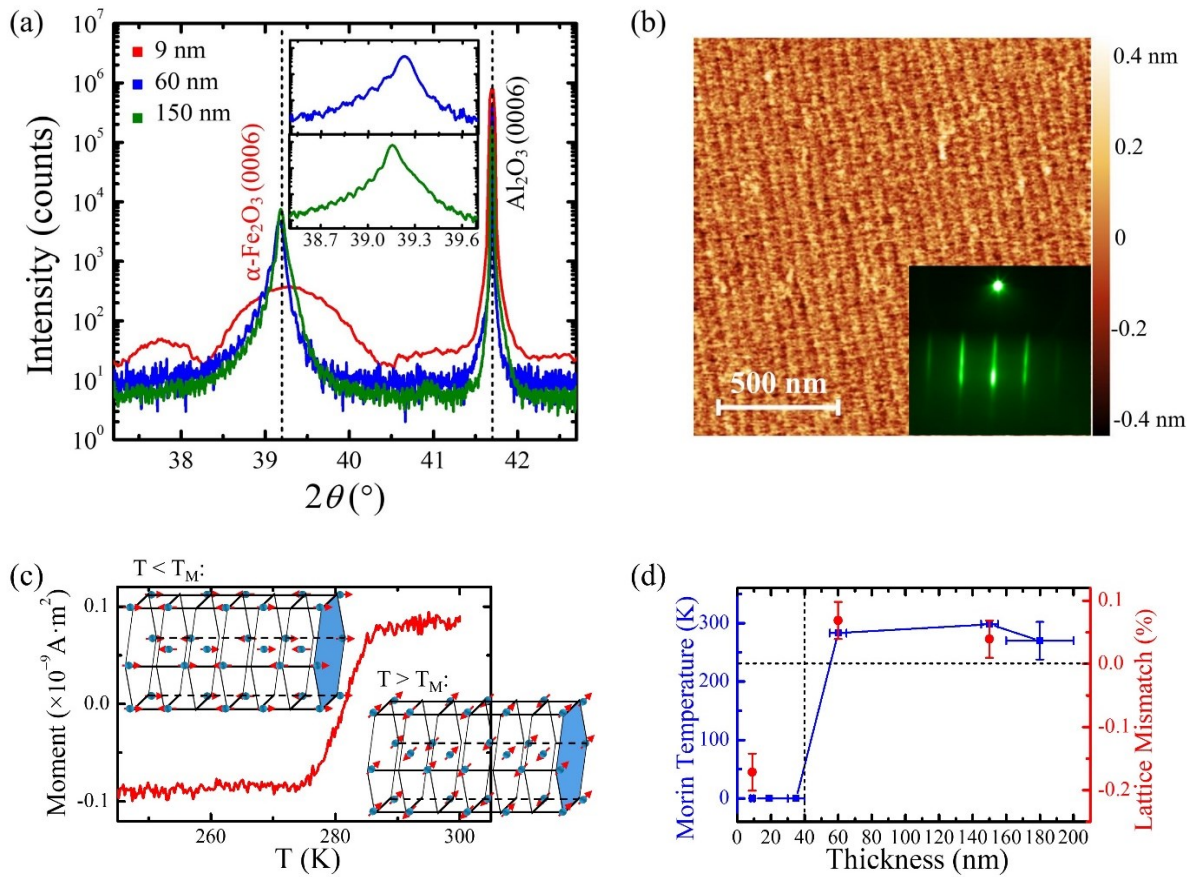

**Figure 1. Hematite α-Fe₂O₃ (0001) thin film characterization and Morin transition**. (a) XRD spectra of α-Fe$_2$O$_3$ thin films of different thicknesses (~ 9 nm, ~ 60 nm, and ~ 150 nm) grown on Al$_2$O$_3$ (0001) substrate. Dashed lines refer to the peak positions of bulk α-Fe$_2$O$_3$ and Al$_2$O$_3$ (0006), respectively. The insets are magnified views of the (0006) peaks, showing clear Kiessig fringes. (b) Atomic force microscope image of a 10 nm hematite (0001) thin film with step edges and an RMS roughness of around 0.1 nm. The inset shows the RHEED pattern of α-Fe$_2$O$_3$ thin film after growth and in-situ cooling, with electron beam applied perpendicular to the [11$\bar{2}$0] direction. (c) SQUID measurements under zero-field cooled conditions reveal a magnetization jump at the Morin transition in the hematite thin film (~ 180 nm). (d) Summary of the Morin temperature of thin films with different thicknesses and lattice mismatch with respect to the bulk α-Fe$_2$O$_3$ along out-of-plane direction, revealing a critical thickness and out-of-plane tensile strain for Morin transition.

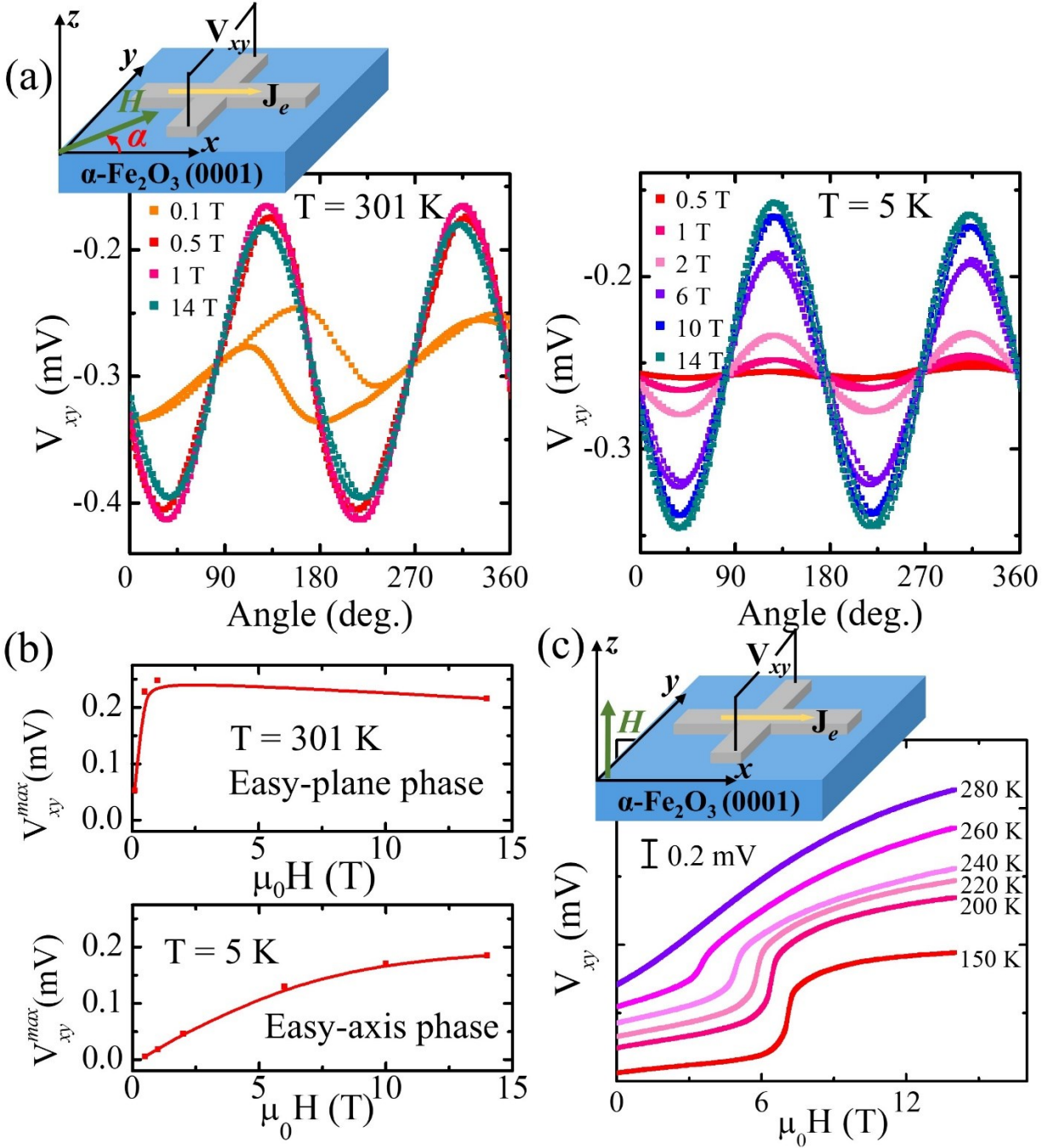

**Figure 2. Transport properties of α-Fe$_2$O$_3$(60nm)/Pt heterostructure.** (a) In-plane angular dependence of the transverse voltage V$_{xy}$ below (T = 5 K) and above (T = 301 K) the Morin temperature under different magnetic fields. (b) Summary of the field dependence of V$_{xy}$ responses in the easy-axis and easy-pane phases. (c) Field sweeps of the transverse voltage with magnetic field along the c-axis.

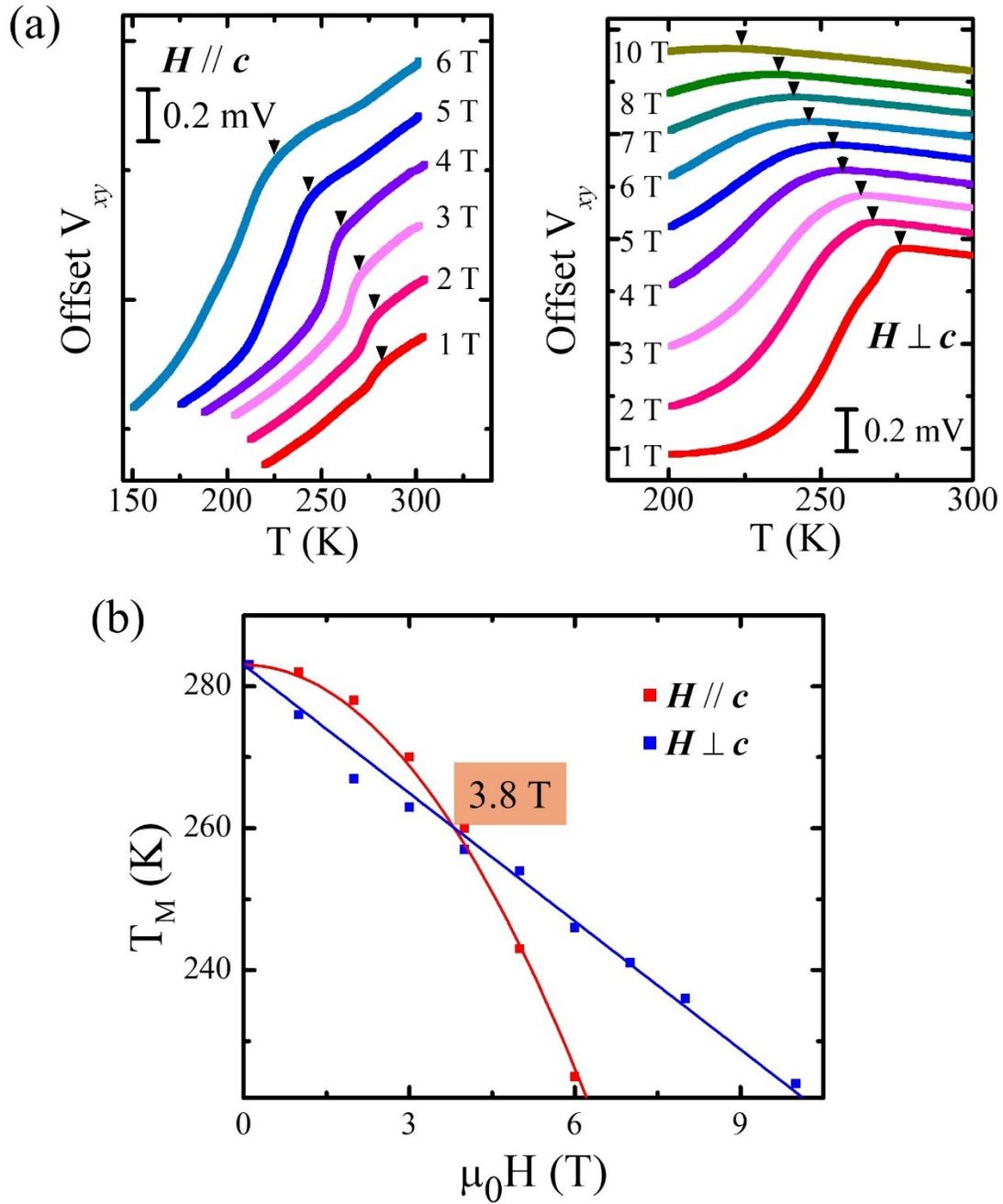

**Figure 3. Anisotropic Morin transition responses to external magnetic fields in hematite thin film.** (a) Temperature dependence of V$_{xy}$ with **H** // **c** (upper) and **H** ⊥ **c** (lower). Results are offset vertically for better clarity and labeled as Offset V$_{xy}$. (b) Field dependence of $T_M$ in two orthogonal field directions extracted from transport measurements. Two curves cross around $\mu_0 H$ ~ 3.8 T.

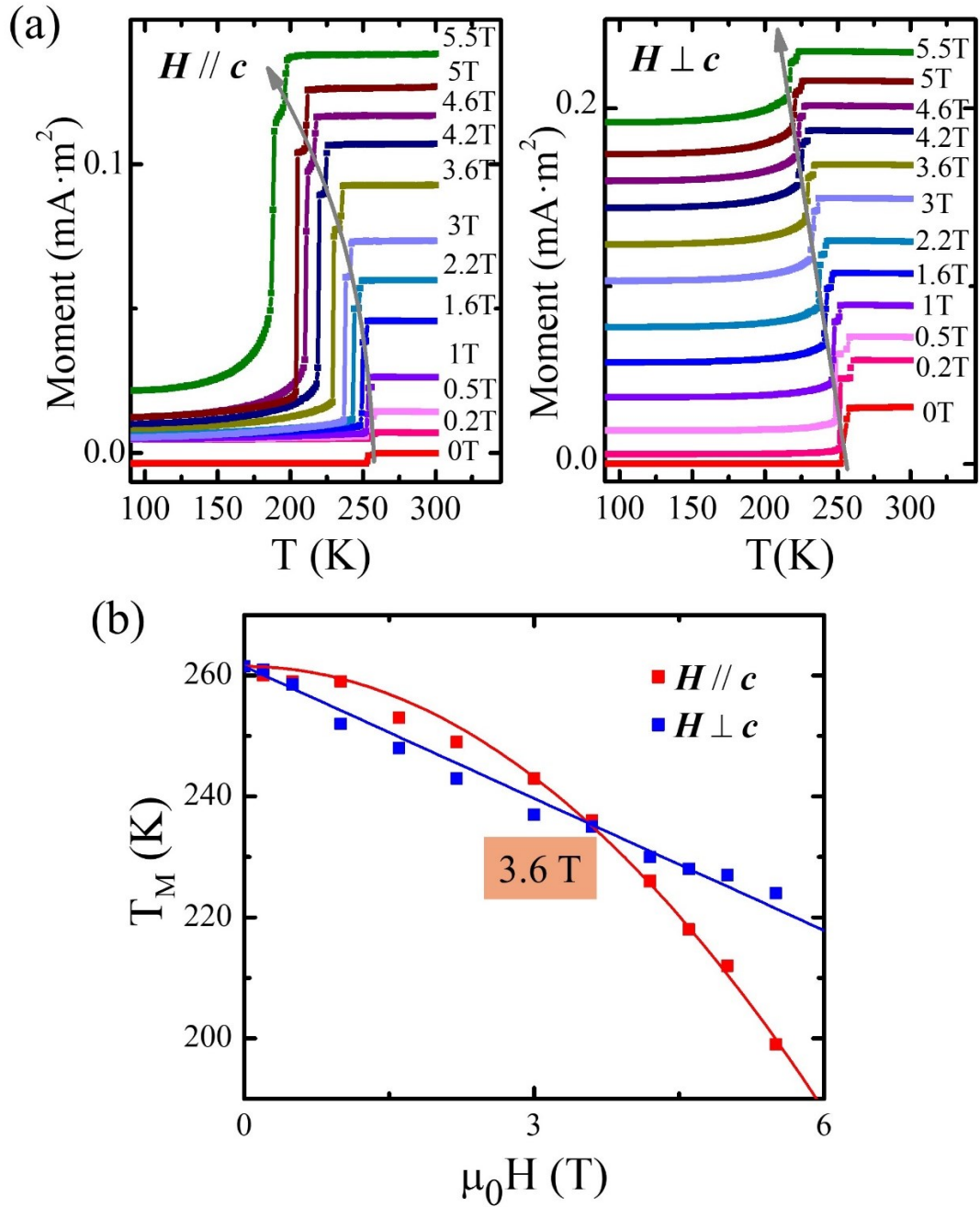

**Figure 4. Anisotropic Morin transition responses to external magnetic fields in hematite bulk crystal.** (a) Temperature dependence of bulk magnetization with $H \parallel c$ (upper) and $H \perp c$ (lower). (b) Field dependence of $T_M$ in two orthogonal field directions extracted from the magnetometry measurements. Two curves cross around $\mu_0 H$ ~3.6 T.

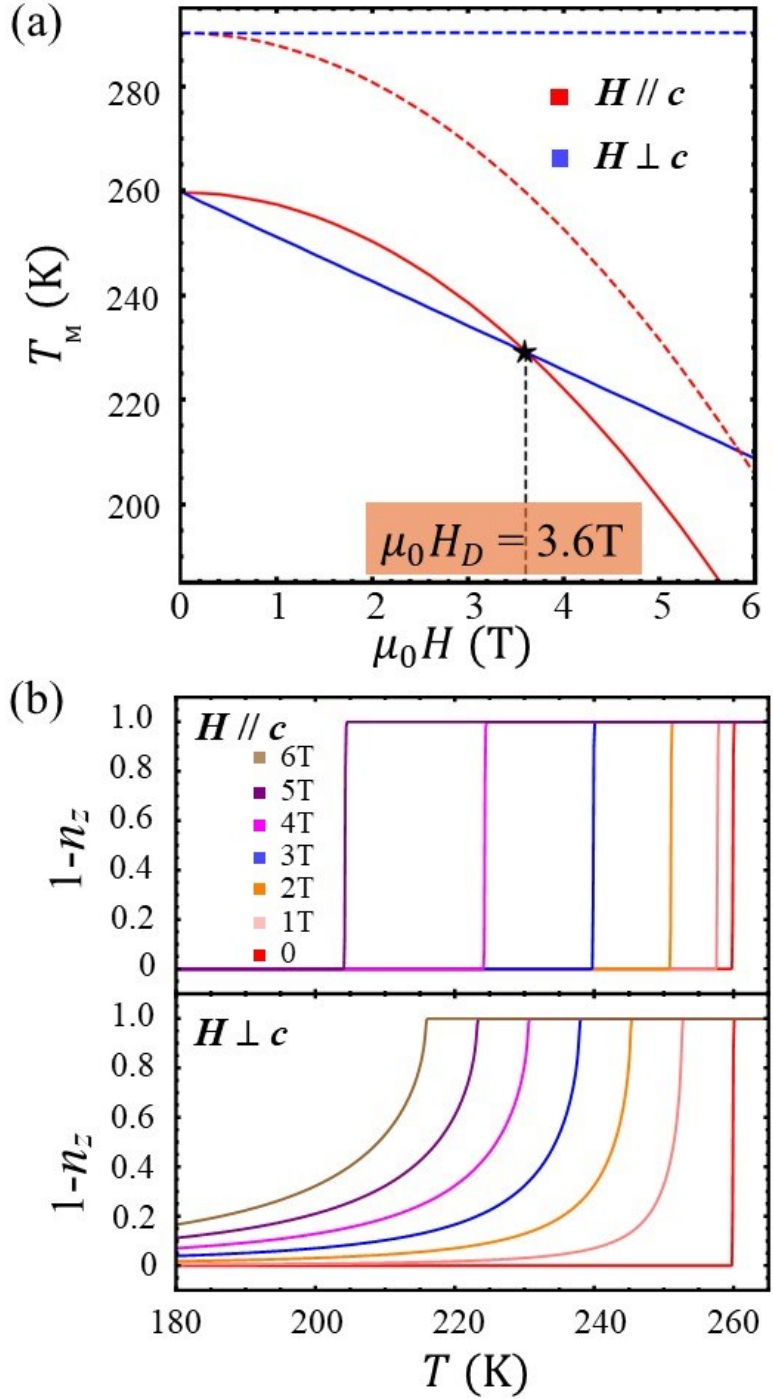

**Figure 5. LLG simulation results.** (a) Simulation of the field dependence of $T_M$. Dashed (solid) lines are results without (with) the DM interaction. The star marks the crossing point between two simulated curves for two field directions which determines the DM interaction strength. (b) Simulation of temperature dependence of $1 - n_x$ under different field strength for $H // c$ (upper) and $H \perp c$ (lower).

# Reference

[1] V. Baltz, A. Manchon, M. Tsoi, T. Moriyama, T. Ono, and Y. Tserkovnyak, Antiferromagnetic spintronics, Rev. Mod. Phys. 90, 015005 (2018).

[2] T. Jungwirth, X. Marti, P. Wadley, and J. Wunderlich, Antiferromagnetic spintronics, Nat. Nanotechnol. 11, 231 (2016).

[3] L. Šmejkal, Y. Mokrousov, B. Yan, and A. H. MacDonald, Topological antiferromagnetic spintronics, Nat. Phys. 14, 242 (2018).

[4] L. Šmejkal, J. Sinova, and T. Jungwirth, Emerging research landscape of altermagnetism, Phys. Rev. X 12, 040501 (2022).

[5] J. Han, R. Cheng, L. Liu, H. Ohno, and S. Fukami, Coherent antiferromagnetic spintronics, Nat. Mater. 22, 684 (2023).

[6] F. J. Morin, Magnetic susceptibility of α-$Fe_2O_3$ and α-$Fe_2O_3$ with added titanium, Phys. Rev. 78, 819 (1950).

[7] I. Dzyaloshinsky, A thermodynamic theory of "weak" ferromagnetism of antiferromagnetics, J. Phys. Chem. Solids 4, 241 (1958).

[8] T. Moriya, Anisotropic superexchange interaction and weak ferromagnetism, Phys. Rev. 120, 91 (1960).

[9] H. Thoma, V. Hutanu, H. Deng, V. E. Dmitrienko, P. J. Brown, A. Gukasov, G. Roth, and M. Angst, Revealing the absolute direction of the Dzyaloshinskii-Moriya interaction in prototypical weak ferromagnets by polarized neutrons, Phys. Rev. X 11, 011060 (2021).

[10] P. Zhang, J. Finley, T. Safi, and L. Liu, Quantitative study on current-induced effect in an antiferromagnet Insulator/Pt bilayer film, Phys. Rev. Lett. 123, 247206 (2019).

[11] Y. Cheng, S. Yu, M. Zhu, J. Hwang, and F. Yang, Electrical switching of tristate antiferromagnetic Néel order in α-$Fe_2O_3$ epitaxial films, Phys. Rev. Lett. 124, 027202 (2020).

[12] X. Chen et al., Control of spin current and antiferromagnetic moments via topological surface state, Nat. Electron. 5, 574 (2022).

[13] P. Zhang, C.-T. Chou, H. Yun, B. C. McGoldrick, J. T. Hou, K. A. Mkhoyan, and L. Liu, Control of Néel vector with spin-orbit torques in an antiferromagnetic insulator with tilted easy plane, Phys. Rev. Lett. 129, 017203 (2022).

[14] R. Lebrun, A. Ross, S. A. Bender, A. Qaiumzadeh, L. Baldrati, J. Cramer, A. Brataas, R. A. Duine, and M. Kläui, Tunable long-distance spin transport in a crystalline antiferromagnetic iron oxide, Nature 561, 222 (2018).

[15] J. Han, P. Zhang, Z. Bi, Y. Fan, T. S. Safi, J. Xiang, J. Finley, L. Fu, R. Cheng, and L. Liu, Birefringence-like spin transport via linearly polarized antiferromagnetic magnons, Nat. Nanotechnol. 15, 563 (2020).

[16] R. Lebrun, A. Ross, O. Gomonay, V. Baltz, U. Ebels, A.-L. Barra, A. Qaiumzadeh, A. Brataas, J. Sinova, and M. Kläui, Long-distance spin-transport across the Morin phase transition up to room temperature in ultra-low damping single crystals of the antiferromagnet α-$Fe_2O_3$, Nat. Commun. 11, 6332 (2020).

[17] J. Gückelhorn et al., Observation of the nonreciprocal magnon Hanle effect, Phys. Rev. Lett. 130, 216703 (2023).

[18] I. Boventer, H. T. Simensen, A. Anane, M. Kläui, A. Brataas, and R. Lebrun, Room-temperature antiferromagnetic resonance and inverse spin-Hall voltage in canted antiferromagnets, Phys. Rev. Lett. 126, 187201 (2021).

[19] H. Wang, Y. Xiao, M. Guo, E. Lee-Wong, G. Q. Yan, R. Cheng, and C. R. Du, Spin pumping of an easy-plane antiferromagnet enhanced by Dzyaloshinskii–Moriya interaction, Phys. Rev. Lett. 127, 117202 (2021).

[20] S. Foner and S. J. Williamson, Low-temperature antiferromagnetic resonance in α-$Fe_2O_3$, J. Appl. Phys. 36, 1154 (1965).

[21] V. E. Dmitrienko, E. N. Ovchinnikova, S. P. Collins, G. Nisbet, G. Beutier, Y. O. Kvashnin, V. V. Mazurenko, A. I. Lichtenstein, and M. I. Katsnelson, Measuring the Dzyaloshinskii–Moriya interaction in a weak ferromagnet, Nat. Phys. 10, 202 (2014).

[22] S. Park, H. Jang, J.-Y. Kim, B.-G. Park, T.-Y. Koo, and J.-H. Park, Strain control of Morin temperature in epitaxial α-$Fe_2O_3$ (0001) film, Europhys. Lett. 103, 27007 (2013).

[23] A. Ross, R. Lebrun, C. Ulloa, D. A. Grave, A. Kay, L. Baldrati, F. Kronast, S. Valencia, A. Rothschild, and M. Kläui, Structural sensitivity of the spin Hall magnetoresistance in antiferromagnetic thin films, Phys. Rev. B 102, 094415 (2020).

[24] H. Jani et al., Antiferromagnetic half-skyrmions and bimerons at room temperature, Nature 590, 74 (2021).

[25] H. Jani et al., Reversible hydrogen control of antiferromagnetic anisotropy in α-$Fe_2O_3$, Nat. Commun. 12, 1668 (2021).

[26] T. D. Wang, A. C. Basaran, R. El Hage, J. Li, H. Navarro, F. E. Torres, O. R. de la Fuente, and I. K. Schuller, Coercivity enhancement in hematite/permalloy heterostructures across the Morin transition, J. Magn. Magn. Mater. 597, 172024 (2024).

[27] D. Kan, T. Moriyama, R. Aso, S. Horai, and Y. Shimakawa, Triaxial magnetic anisotropy and Morin transition in α-$Fe_2O_3$ epitaxial films characterized by spin Hall magnetoresistance, Appl. Phys. Lett. 120, (2022).

[28] M. Scheufele, J. Gückelhorn, M. Opel, A. Kamra, H. Huebl, R. Gross, S. Geprägs, and M. Althammer, Impact of growth conditions on magnetic anisotropy and magnon Hanle effect in α-$Fe_2O_3$, APL Mater. 11, (2023).

[29] M. A. Tanaka, K. Yokoyama, A. Furuta, K. Fujii, and K. Mibu, Thickness dependence of Morin transition of Ru-doped α-$Fe_2O_3$ films detected by spin Hall magnetoresistance measurements, J Appl. Phys. 135, (2024).

[30] H. Nakayama et al., Spin Hall magnetoresistance induced by a nonequilibrium proximity effect, Phys. Rev. Lett. 110, 206601 (2013).

[31] T. Kosub, M. Kopte, F. Radu, O. G. Schmidt, and D. Makarov, All-electric access to the magnetic-field-invariant magnetization of antiferromagnets, Phys. Rev. Lett. 115, 097201 (2015).

[32] Y. Cheng, S. Yu, A. S. Ahmed, M. Zhu, Y. Rao, M. Ghazisaeidi, J. Hwang, and F. Yang, Anisotropic magnetoresistance and nontrivial spin Hall magnetoresistance in Pt/α-$Fe_2O_3$ bilayers, Phys. Rev. B 100, 220408 (2019).

[33] T. Moriyama, Y. Shiratsuchi, T. Iino, H. Aono, M. Suzuki, T. Nakamura, Y. Kotani, R. Nakatani, K. Nakamura, and T. Ono, Giant anomalous Hall conductivity at the Pt/$Cr_2O_3$ interface, Phys. Rev. Appl. 13, 034052 (2020).

[34] J. Fischer et al., Spin Hall magnetoresistance in antiferromagnet/heavy-metal heterostructures, Phys. Rev. B 97, 014417 (2018).

[35] L. Baldrati et al., Full angular dependence of the spin Hall and ordinary magnetoresistance in epitaxial antiferromagnetic NiO(001)/Pt thin films, Phys. Rev. B 98, 024422 (2018).

[36] R. Lebrun et al., Anisotropies and magnetic phase transitions in insulating antiferromagnets determined by a Spin-Hall magnetoresistance probe, Commun. Phys. 2, 50 (2019).

[37] T. Z. Zhang, K. K. Meng, X. Shi, Y. B. Deng, L. B. Zhu, X. G. Xu, and Y. Jiang, Spin Hall magnetoresistance in antiferromagnetic α-$Fe_2O_3$/Pt bilayers: Modulation from interface magnetic state, Appl. Phys. Lett. 121, 262404 (2022).

[38] J. Fischer, M. Althammer, N. Vlietstra, H. Huebl, S. T. B. Goennenwein, R. Gross, S. Geprägs, and M. Opel, Large spin Hall magnetoresistance in antiferromagnetic α-$Fe_2O_3$/Pt heterostructures, Phys. Rev. Appl. 13, 014019 (2020).

[39] A. Wittmann et al., Role of substrate clamping on anisotropy and domain structure in the canted antiferromagnet α-Fe2O3, Phys. Rev. B 106, 224419 (2022).

[40] J. H. Han, C. Song, F. Li, Y. Y. Wang, G. Y. Wang, Q. H. Yang, and F. Pan, Antiferromagnet-controlled spin current transport in SrMnO3/Pt hybrids, Phys. Rev. B 90, 144431 (2014).

[41] Y. Shapira, Ultrasonic behavior near the spin-flop transitions of hematite, Phys. Rev. 184, 589 (1969).

[42] N. Papanicolaou, Dynamics of domain walls in weak ferromagnets, Phys. Rev. B 55, 12290 (1997).

[43] A. N. Bogdanov, U. K. Rößler, M. Wolf, and K.-H. Müller, Magnetic structures and reorientation transitions in noncentrosymmetric uniaxial antiferromagnets, Phys. Rev. B 66, 214410 (2002).

[44] H. Zhang and R. Cheng, Theory of harmonic Hall responses of spin-torque driven antiferromagnets, J. Magn. Magn. Mater. 556, 169362 (2022).

[45] P. J. Flanders and S. Shtrikman, Magnetic field induced antiferromagnetic to weakferromagnetic transitions in hematite, Solid State Commun. 3, 285 (1965).